# Transcoder for the spatial and temporal modes of a photon

Shuai Shi,[1,2] Dong-Sheng Ding,[1,2,3] Zhi-Yuan Zhou,[1,2] Yan Li,[1,2] Wei Zhang,[1,2] Bao-Sen Shi,[1,2,4] and Guang-Can Guo[1,2]

[1]Key Laboratory of Quantum Information, University of Science and Technology of China, Hefei, Anhui 230026, China
[2]Synergetic Innovation Center of Quantum Information & Quantum Physics, University of Science and Technology of China, Hefei, Anhui 230026, China
[3]dds@ustc.edu.cn
[4]drshi@ustc.edu.cn

Encoding information in light with orbital angular momentum (OAM) enables networks to increase channel capacity significantly. However, light in only the fundamental Gaussian mode is suitable for fibre transmission, and not higher order Laguerre Gaussian modes, which carry OAM. Therefore, building a bridge to interface light with OAM and Gaussian mode time-binning is crucially important. Here, we report the realization of a photonic space-time transcoder, by which light with an arbitrary OAM superposition is experimentally converted into a time-bin Gaussian pulse, and vice versa. Furthermore, we clearly demonstrate that coherence is well conserved and there is no cross-talk between orthogonal modes. This photonic device is simple and can be built with scalable architecture. Our experimental demonstration paves the way towards a mixed optical communication in free-space and optical fibre.

**1. Introduction**

Structured vortex beams have many interesting physical properties [1, 2]. Allen et al. first recognized that light beams with phase front $\exp(il\alpha)$ carry an orbital angular momentum (OAM) of $l\hbar$ per photon, $l$ being a topological charge and $\alpha$ representing an azimuthal angle [3]. The unique properties of helically phased light beams are useful in many fields, such as optical tweezers, astronomy, and optical information processing [4-6]. Because of the inherent infinity of OAM states, photons have the potential to encode and carry an unlimited amount of information, enabling the realization of high channel-capacity communications [7]. Recent progress based on photonic OAM and polarization multiplexing for high-speed data transmission clearly shows the advantages of OAM beams in classical optical communications [8]. In quantum information science, besides high-capacity, light encoded in OAM degree of freedom enables alignment-free quantum key-distribution [9]. However, light with OAM has difficulty propagating through commercial optical fibres; it is more suited for transmission in free space.

As a basic scheme for information encoding in communications, time-bin encoding is a technique widely used in classical information processing and quantum information science, where the information is encoded in different time bins of light in the Gaussian mode. It is therefore more suited for long-distance transmission in fibre because of its robustness against optical decoherence [10]. In addition, the time-bin degree of freedom is theoretically unbounded, and therefore constitutes a high-dimensional space like that for the OAM degree of freedom.

A global communication network should be a combination of two sub-networks, one in free space and the other a fibre system. Therefore, converting information freely between OAM and time-bin degrees of freedom, which span spatial and time domains, respectively, is very appealing. In this work, we report on the realization of a space-time transcoder, by which information encoded in the OAM degree of freedom is converted into information encoded in the time domain, and vice-versa. Furthermore, we demonstrate that light coherence is also conserved during the conversion. Additionally, the conversion efficiency in our scheme can in principle reach 100%. Some other schemes can detect several different $l$ states in the time domain [11, 12], but with a theoretical efficiency that cannot exceed the reciprocal of the number $l$. In these schemes, they divided the input pulse into several parts, and then performed projective measurement on them respectively, so the sum of conversion efficiencies for different modes can't be more than one; meanwhile, conservation of coherence has not been proved. The sum of conversion efficiencies of different modes in our scheme is larger than one, it is greater than one shows that our scheme is essentially different. The OAM states can also be measured efficiently by transforming OAM into linear momentum [13, 14], while the sorted light beams can't be effectively coupled into a single mode fiber.

This transcoder has important applications in connecting traditional fibre-based communication systems with satellite-based networks, and also in quantum key distribution without requiring reference frame alignment [9, 15]. The demonstrated space-time device is very simple and also scalable to high OAM states, and appears promising in classical mixed communications with photons exhibiting different degrees of freedom.

**2. Spatial to temporal conversion:**

Our space-time photonic transcoder consists of an optical cavity and an optical loop. The mode selection property of the optical cavity, for which light of different OAM values corresponds to different eigen-frequencies within the cavity, ensures a high extinction ratio between different OAM states of light (See Appendix). When a cavity is locked to the fundamental Gaussian mode, it allows light with zero OAM to be transmitted and reflects all other light with high effectiveness. With a vortex phase plate (VPP) inside, the optical loop is used to trap light pulses and decrease the OAM value of light by 1 per loop. A pulse with OAM $l$ cannot pass through the cavity until its OAM is reduced to zero after $l$ loops. Thus, an optical pulse with different OAM will exit the transcoder at different times $t_0+T*l$, with T the round trip propagation time in the loop. An optical cavity is usually used for purifying a Gaussian laser beam [16], so it could vastly reduce cross-talk between different OAM light. The conversion is expressed as follows:

$$\sum_l \alpha_l |l\rangle \to \sum_l \frac{\beta \alpha_l}{\gamma^l} |t_0 + T*l\rangle \quad (1)$$

where $\alpha_l$ the coefficient associated with OAM value $l$, $\gamma$ the circulation loss of the optical loop, $t_0$ the output time of a Gaussian input pulse, and $\beta$ a normalization coefficient.

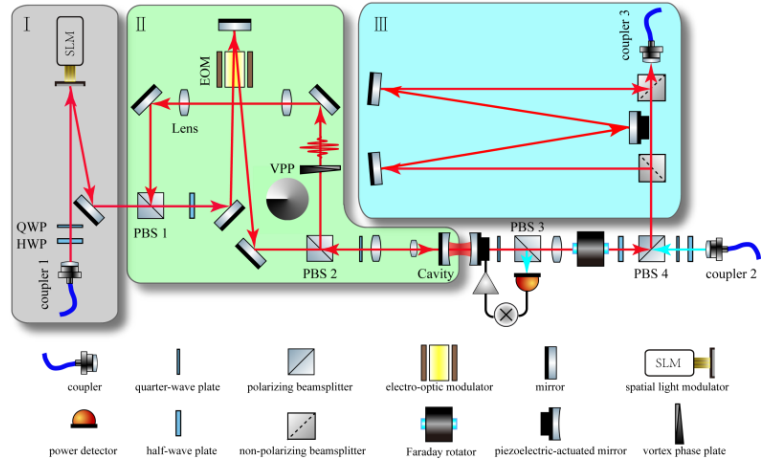

Fig. 1. Schematic of spatial to temporal conversion. In Part I, A Gaussian pulse from coupler 1 carries OAM $l$ after it is diffracted off the SLM. Part II consists of an optical cavity and an optical loop. The EOM is triggered to change the polarization of the pulse just after it double-passes through for the first time. With a vortex phase plate (VPP) inside, the optical loop is used to trap light pulses and decrease the OAM value of light by 1 per loop. The cavity is locked to the fundamental Gaussian mode, it allows light with zero OAM to be transmitted and reflects all other light with high effectiveness. Part III is used to check whether coherence is maintained in the conversion.

In our experiment, the initial continuous wave (CW) laser beam is from a Ti: sapphire laser (Coherent MBR 110) centred at wavelength 795 nm. We use two Pockels cells (Cleveland Crystals Inc., IPD-2545) as fast optical switches to chop the CW laser beam into pulses of 5-ns duration at a 1-kHz repetition rate. In the experimental setup (Fig. 1), the pulsed Gaussian laser beam from coupler 1 (see Part I) is diffracted off a computer-generated fork-diffraction pattern on the spatial light modulator (SLM; Holoeye LETO LCoS). The fork dislocation in the patterns introduces helical phase fronts ($\exp(il\alpha)$) to the first-order diffracted pulse, thereby imparting an OAM of $l$. Part II is the core of our scheme. We use the diffracted pulse from Part I to create a state $|H,l\rangle$, with OAM $l$ and horizontal (H) polarization. The polarizing beam splitter (PBS) is assumed to transmit H-polarized pulses and reflect vertical (V)-polarized pulses. The pulse can then pass through the PBS 1 because the electro-optic modulator (EOM) is triggered just after the pulse double-passes through it for the first time. Hence, the pulse is still in state $|H,l\rangle$, and goes through PBS 2 to the optical cavity. The optical cavity consists of two mirrors each of radius 50 mm and spaced 10 mm apart. As the cavity is locked to the LG$_{00}$ mode, the pulse passes through efficiently if and only if $l=0$, otherwise it is reflected almost totally. The experimentally measured transmissivity for the Gaussian mode is 90% and the reflectivity for nonzero OAM modes is higher than 90%. Even though part of the Gaussian mode is reflected back, high extinction ratios between different OAM states can still be achieved, because the reflected part returns with nonzero OAM, which cannot pass through the cavity. The reflected pulse becomes a state $|V,l\rangle$ after it double-passes the quarter-wave plate (QWP). Next, it is reflected onto the vortex phase plate (VPP) by PBS 2. The VPP introduces a phase exp(-iα) to the pulse after it passing through, so its state becomes $|V,l-1\rangle$. A 4-f system is used to erase detrimental effects to beam quality after a round trip. The EOM changes the polarization of the pulse when it double-passes for the second time, and the state now becomes $|H,l-1\rangle$. The pulse returns repeatedly to the optical cavity with lower OAM, until its OAM becomes zero, whereupon it passes through the cavity. As the total number of reflections in the optical loop is even, we can ignore the sign inversion at each reflection. The round trip propagation time of the optical loop is $T = 11$ ns. In this way, the optical loop converts the pulse with OAM $l$ into a Gaussian pulse at a specific time correlated with the OAM value of the input pulse. Assuming that a Gaussian pulse exits the optical loop at time $t_0$, a pulse with OAM $l$ needs to undergo multiple passages through the VPP and come out at time $t_0+T*l$.

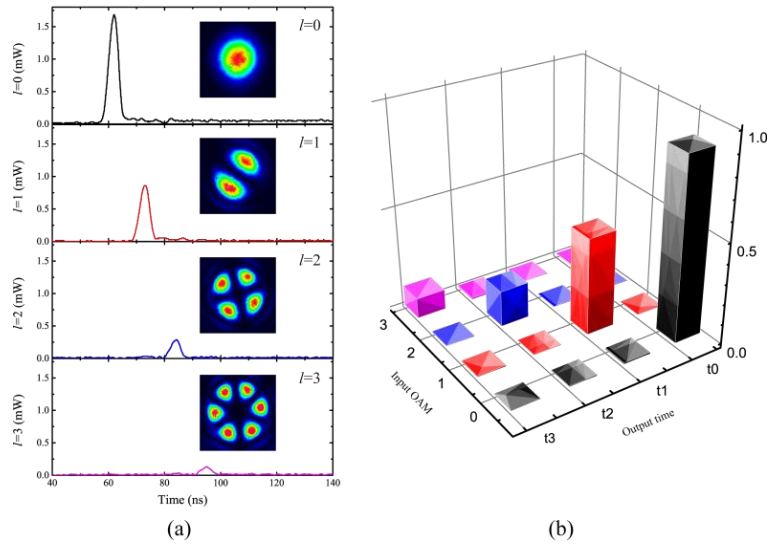

Fig. 2. Experimental results. (a) Waveforms showing the different time bins generated from the input pulses with different OAM; the time interval between adjacent bins is 11 ns. Insets: interference patterns of the input light beams with different OAM values, generated by Part I. (b) The conversion efficiencies of OAM to time-bin conversion.

To confirm the OAM of the light beam generated from Part I, we brought the light beam and its own mirror image to interfere. A Mach-Zehnder (M-Z) interferometer with even and odd numbers of reflections in the different arms can achieve this goal, because the sign of the OAM is inverted at each reflection. For light with OAM $l$, the interference pattern has $2l$ radial fringes. The insets in Fig. 2(a) show the interference patterns for light beams of different OAM values, generated by Part I. Pulses with different OAM values are input into the transcoder at the same time and pass through the cavity at different times. Next, they are coupled into a single mode fibre using coupler 3. Finally, the output from coupler 3 is detected by a fast photodiode (Thorlabs SUV-7) connected to a 500-MHz oscilloscope for time-domain detection. The waveforms in Fig. 2(a) show the converted Gaussian pulses at different times. From these waveforms, we can clearly see the relationship between the output time $t_l$ of the Gaussian pulse and the OAM $l$ of the input pulse: $t_l = t_0 + T * l$.

In our experiment, the time interval $T$ between the generated adjacent bins is 11 ns. The intensity attenuation of the pulse is mainly caused by optical loss in the optical loop, which can be improved using high-quality optical components, and thus it can work for higher OAM orders. In our experiment the average transmissivity of the uncoated EOM and VPP is 90%, the average transmissivity and reflectivity of the rest components is 99%, so the optical components limit the efficiency to 60.2%. The rest of the loss is caused by the mismatch between the input pulse and the optical cavity after the optical loop. The $\gamma$ for the current setup is 2.06. The conversion efficiency can reach 82% if the average transmissivity and reflectivity of the optical components can be improved to

99.5% and the mode coupling efficiency of the pulse to cavity is improved to 90% after the optical loop. Under these conditions, the transcoder can convert more than 10 OAM modes.

| Cross-talk (dB) | $l = 0$ | $l = 1$ | $l = 2$ | $l = 3$ |
|---|---|---|---|---|
| $t(l) - T$ | * | -25.2 | -18.45 | -11.4 |
| $t(l) + T$ | -21.3 | -24.4 | -19.2 | * |

Table 1. The nearest-neighbour cross-talk among the four OAM states, $t(l)\pm T$ corresponds to the output obtained before and later than the correct time.

To check whether there is cross-talk in the process of conversion, we integrated the intensity under each peak versus the input OAM state $l$ and the output time bin $t_l$. The conversion efficiencies are shown in Fig. 2(b). Cross-talk is defined as the ratio between the neighbouring incorrect detections and the correct detection. From the normalized matrix shown in Fig. 2(b), we obtain the nearest-neighbour cross-talk among the four OAM states (Table 1.), the average cross-talk is –20 dB. Low cross-talk means low bit error rate; this property can also be used to measure the OAM spectrum.

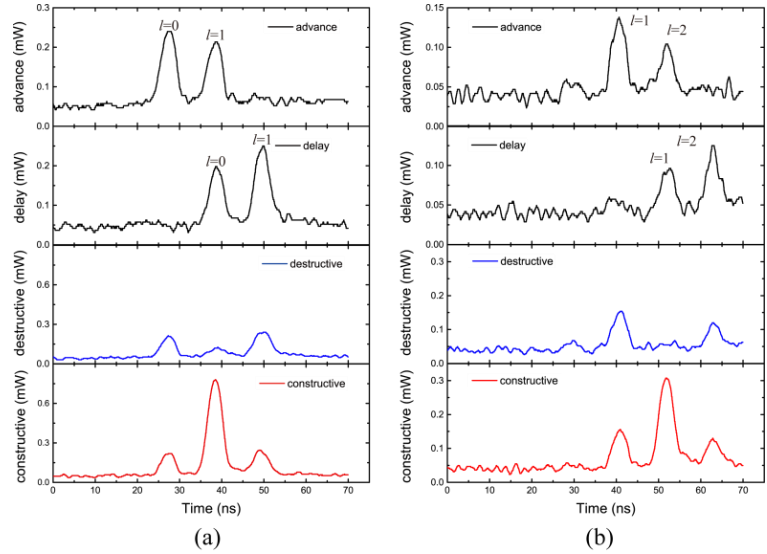

Fig. 3. Results to check coherence during conversion. The advance and delay curves represent the detected pulses from the short arm and long arm, respectively, of the M–Z interferometer in Part III of the transcoder (Fig. 1). Pulses from the different arms have components overlapping in time, which interfere with each other when they come together. (a) Interference waveforms for a pulse in a superposition of OAM states $l=0$ and $l=1$. (b) The interference waveforms for a pulse in a superposition of OAM states $l=1$ and $l=2$.

We used an unbalanced M-Z interferometer in creating interference between adjacent pulses to verify whether coherence is maintained in the conversion. One arm of the interferometer is 3.3 m longer than the other. A pulse formed by the superposition of OAM states $l=0$ and $l=1$: $\sin\beta|0\rangle + \cos\beta|1\rangle$ is converted into a time-domain superposition of states: $\gamma\cdot\sin\beta|t_0\rangle + \frac{\cos\beta}{\gamma}|t_0+T\rangle$, where $\gamma$ is related to the different optical loss in the two components. When the pulse output from the M-Z interferometer, the two components interfere with each other in Part III (Fig. 1). In Fig. 3(a), the advance and delay curves represent the detected pulses from the short arm and long arm, respectively. In the experiment, we ensure the two components have approximately equal strength by adjusting the proportion of the different components and the optical losses in one arm of the interferometer. Because of fluctuations in the arm-length difference of the interferometer, we observed the pattern to change slowly between constructive and destructive interference [blue and red curves, respectively, in Fig. 3(a)]. The interference visibility reaches 0.795 after background electrical noise is corrected, demonstrating that coherence is conserved in the conversion. Further tests with the superposition of OAM $l=1$ and $l=2$ states comes to the same conclusion, and the interference visibility is 0.806 [see Fig. 3(b)]. There are two major reasons leading to the visibility limited to 80%, one is decoherence during the conversion, the other one is the intensity jitter of the pulses caused by the instability of the setup. Because coherence during the conversion is well conserved, the transcoder is suitable for any coherent optical communication system.

### 3. Temporal to spatial conversion:

Interestingly, when the same setup functions under inverse cycling, pulses input at specific times can be transformed into corresponding OAM modes. In this way, the functions of the optical system are changed. Part III is used to generate two coherence pulses with a time interval of 11 ns. Part II introduces a phase front $\exp(il\alpha)$ to the pulse input at time $t_0-T*l$. Finally, Part I performs a projective measurement for the spatial structure of the generated pulse. The conversion could be described as follows:

$$\sum_l \alpha_l |t_0 - T*l\rangle \rightarrow \sum_l \frac{\beta \alpha_l}{\gamma^l} |l\rangle \quad (2)$$

where $\alpha_l$ is the proportion of the component input at time $t_0-T*l$, $\gamma$ the circulation loss of the optical loop, $t_0$ the input time for a Gaussian output pulse, and $\beta$ the normalization coefficient.

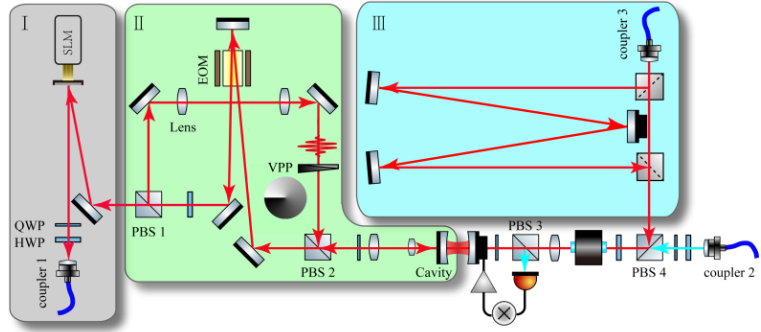

Fig. 4. Schematic of temporal to spatial conversion. Gaussian pulse from coupler 3 passes through the cavity at time $t_0-T*l$, The pulse returns repeatedly to the optical cavity with higher OAM, until the EOM is triggered to change the polarization of the pulse at time $t_0$. Then the pulse passes through PBS 1 in state $|H,l\rangle$ at time $t_0$. Finally, Part I performs a projective measurement for the spatial structure of the generated pulse. Part III is used to check whether coherence is maintained in the conversion.

First, we block one arm of the M-Z interferometer in the experimental setup (Fig. 4). Gaussian laser beam from coupler 3 mode-matches with the cavity. The laser beam from coupler 2 is used to lock the cavity onto the Gaussian mode. Because the peak power of the pulse is much stronger than the laser beam from coupler 2, the influence of the lock-cavity laser beam on the pulse is negligible. Gaussian pulse from coupler 3 passes through the cavity at time $t_0-T*l$, the pulse is still in state $|H,0\rangle$ after passing through PBS 2 and EOM, and then the half-wave-plate (HWP) near PBS 1 changes the polarization of the pulse, so its state becomes $|V,0\rangle$. Next, the VPP introduces a phase $\exp(i\alpha)$ to the pulse after it passing through along the opposite direction, the state becomes $|V,1\rangle$. The pulse returns repeatedly to the optical cavity with higher OAM, until the EOM is triggered to change the polarization of the pulse at time $t_0$. Then the pulse passes through PBS 1 in state $|H,l\rangle$ at time $t_0$. The output pulse is diffracted off the fork-diffraction patterns on the SLM. The fork-diffraction pattern displayed on the SLM can be used to "flatten" the phase of the component with a specific incident OAM, which can be coupled efficiently to a single-mode fibre. Finally, we use a fast photodiode to measure how much power was carried by the mode we are projecting into.

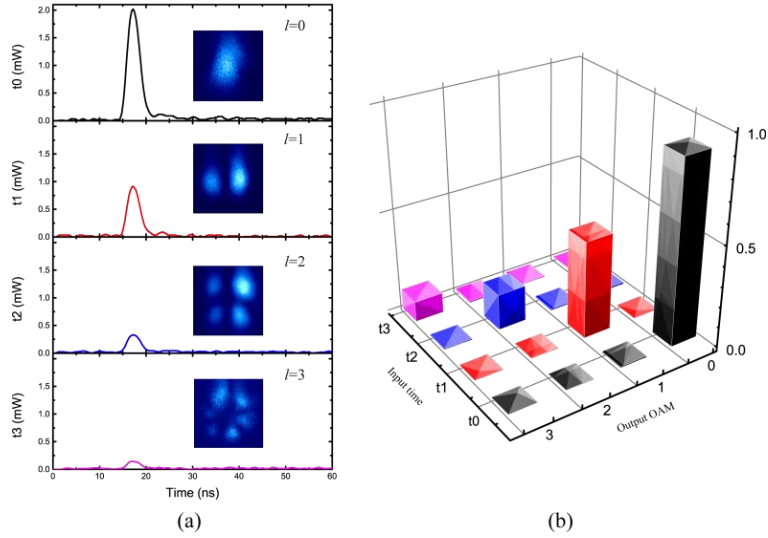

Fig. 5. (a) Projective measurement waveforms for pulses generated with different OAM from pulses input at different times, and the detected pulses in the corresponding OAM mode. Insets show the interference patterns of the generated pulses with their own mirror image. (b) The conversion efficiencies of time-to-OAM conversion.

As shown in Fig. 5(a), pulses input at different time steps $t_i$ can be converted to output pulses at the same time with a specific OAM $l$=i corresponding to the input time. To show the OAM of the generated pulses, we use the M-Z interferometer mentioned before to create interference of the generated pulses with its own mirror image. The insets in Fig. 5(a) show these interference patterns which are detected by a fast-gated intensified CCD camera (ANDOR iStar 334T) with a gating speed of 6 ns. An interference pattern with $2l$ radial fringes clearly indicates that the pulse carried OAM $l$=i.

| Cross-talk (dB) | $t = 0$ | $t = -T$ | $t = -2T$ | $t = -3T$ |
| --- | --- | --- | --- | --- |
| $l(t) - 1$ | * | -31.08 | -14.17 | -11 |
| $l(t) + 1$ | -22.1 | -22.38 | -16.31 | * |

Table 2. The nearest-neighbour cross-talk among the four OAM states, $l(t)\pm 1$ corresponds to the output with higher and lower OAM than the correct OAM states.

To check cross-talk in this conversion, we measured the OAM content of the generated pulse for each input time via a projective measurement. The conversion efficiencies are shown in Fig. 5(b). The nearest-neighbour cross-talk among the four OAM states are shown in Table 2., the average cross-talk is –19.5dB. The same setup, when used in reverse, generates pulses with OAM at high frequency. In our experiment, the frequency is 1 kHz, which is much higher than the image frame rate (60 Hz) of spatial light modulator and could be improved easily. Rapid generation of light beams carrying OAM is an enabling technology for classical communication that employ spatial mode encoding [17].

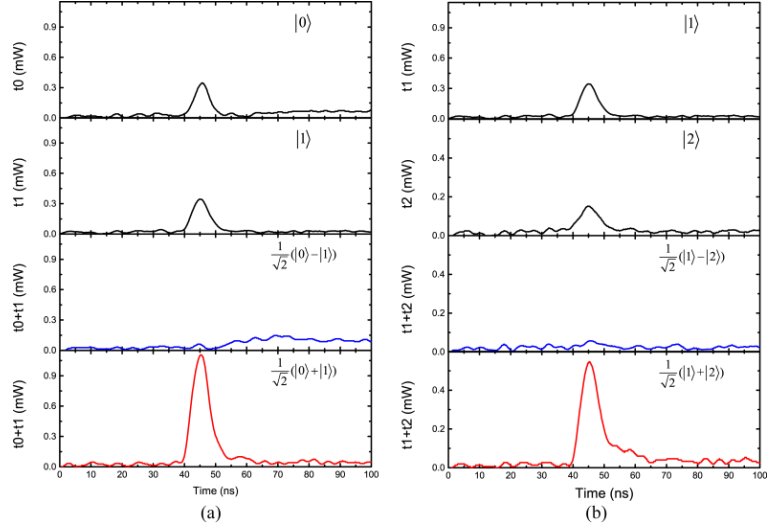

Fig. 6. Curves of the projective measurement result on OAM state $|i\rangle$: (a) Interference result for pulses input at time $t_0$ and $t_1$; (b) Interference result for pulses input at time $t_1$ and $t_2$. (Here ti represents the input time of the pulse.)

To verify whether coherence is maintained in the conversion, the M-Z interferometer in Part III of the transcoder (Fig. 1) is used to prepare a pulse that is a superposition of time bin states $|t_i\rangle$ and $|t_{i+1}\rangle$. At different time bins, the pulse is converted into pulses with different OAM. When they output at the same time, the generated pulse will be in a superposition of OAM states $l=i$ and $l=i+1$. Performing a projective measurement on such superpositions, $\frac{1}{\sqrt{2}}(|i\rangle-|i+1\rangle)$ and $\frac{1}{\sqrt{2}}(|i\rangle+|i+1\rangle)$, we get destructive and constructive interference (Fig. 6). The interference visibility reached 0.92 and 0.792 for i=0 and i=1 respectively. Obviously, the coherence is maintained in the conversion, indicating that the transcoder can prepare pulses that are superpositions of OAM states.

### 4. Discussion

Because optical losses in the imperfect optical loop are significant, we have at present only demonstrated optical conversion with just four OAM modes. In principle, one can convert much more topological charge of OAM modes by improving the efficiencies. Moreover, this photonic device could work with shorter pulses by changing the parameters of the cavity. We can then miniaturize the optical loop, and hence, this device could function in a more scalable architecture.

Our transcoder can be regarded as an OAM sorter or a high frequency OAM generator, it can also be used in studying photonic entanglement within the OAM and time-bin degrees of freedom, and may be useful in studying quantum simulation using OAM modes. In this way, one can easily prepare hybrid entanglements with OAM and time-bin states used in general nonlinear processes such as spontaneously parametric down conversion and spontaneously four-wave mixing.

The present setup can only separately work for positive or negative OAM values. There are two ways to make the setup work for both positive and negative OAM values simultaneously: one is to place a phase plate in front of PBS 1 to increase the OAM value, which only works for a small amount of negative OAM values; the other one is to place a half-wave plate in front of PBS 1. By this way, half of the input pulses are reflected out and into the optical loop via PBS 2, as thus half of the input pulses propagate along the opposite direction in the optical loop, the phase plate will increase the OAM value by 1 per loop. This way can only work with 50% efficiency.

### 5. Conclusion

We have realized the conversion of pulses with spatial information encoded in the OAM degree of freedom into the time domain and vice versa. Furthermore, we have demonstrated the conservation of coherence during the conversion. This space-time transcoder is very simple and scalable for high OAM states. High extinction ratios between different OAM states have been achieved, and the main restriction for high-order OAM conversion stems from the losses of the uncoated components in the optical loop. Our work offers a very promising application to classical mixed communications with photons exhibiting large degrees of freedom.

**Appendix**

Optical Cavity:

The transmission of an optical cavity is

$$T = \frac{1}{1 + \frac{4R}{(1-R)^2}\sin(\frac{2\pi}{c}\nu nd)} \qquad (3)$$

The finesse of the cavity is

$$F = \frac{\pi\sqrt{R}}{1-R} = \frac{\pi\sqrt{0.95}}{1-0.95} \approx 61.24 \qquad (4)$$

where $R$ is the reflectivity of the cavity mirror.

The free spectral range (FSR) is

$$\Delta\nu = \frac{c}{2nd} \approx \frac{3\times10^8 \text{m/s}}{2\times1\times0.01\text{m}} = 15GHz \qquad (5)$$

where $d$ is the distance between the cavity mirrors.

The full-width-at-half-maximum (FWHM) of the transmission peak is

$$FWHM = \frac{\Delta\nu}{F} \approx 245MHZ \qquad (6)$$

The linewidth for a 5-ns laser pulse is 16 MHz, which is less than the FWHM, thereby ensuring transmission of the pulse.

The electric field of the Laguerre-Gaussian (LG) modes in cylindrical coordinates is written [2]

$$LG_{pl} = \sqrt{\frac{2p!}{\pi(p+|l|)!}}\frac{1}{w(z)}(\frac{r\sqrt{2}}{w(z)})^{|l|}\exp(\frac{-r^2}{w^2(z)})L_p^{|l|}(\frac{2r^2}{w^2(z)})\exp(il\alpha)\exp(\frac{ik_0 r^2 z}{2(z^2+z_R^2)}) \\ \exp(-i(2p+|l|+1)\tan^{-1}(\frac{z}{z_R})) \qquad (7)$$

The factor $(2p+|l|+1)\tan^{-1}(\frac{z}{z_R})$ is the Gouy phase, which is responsible for dispersion in the transverse direction. The phase shift determines the different eigenfrequency of the LG modes within a resonator which are expressed as

$$\nu_{l,p,m} = \frac{c}{2nd}(m + \frac{2p+|l|+1}{\pi}\cos^{-1}(\sqrt{g_1 g_2})) \qquad (8)$$

where $l$ and $p$ are transverse quantum numbers, m the longitudinal quantum number, and $g_i = (1-\frac{d}{R_i})$ the stability parameter. In our experiment, $g_i = 0.8$, and hence:

$$\nu_{l,p,m} = \frac{c}{2nd}(m + (2p+|l|+1)\times 0.2048) \qquad (9)$$

Even though the eigenfrequencies of some OAM values are close to those of the Gaussian modes, their frequency difference can be bigger than FWHM for a wide range of OAM values. In our experiment, the range is OAM $l$<201. The experimentally measured reflectivity is greater than 85%, and the transmittivity is 6×10−5 for OAM $l$=5. The value is better for laser beam with other OAM values.

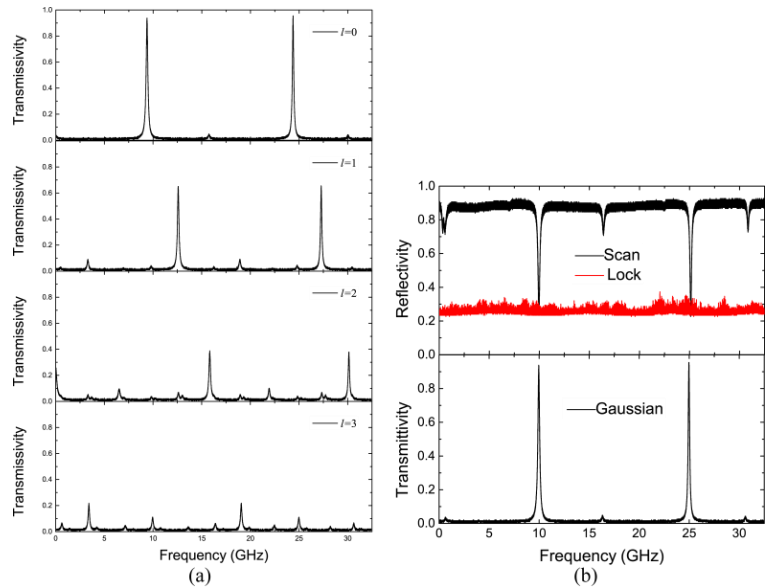

Fig. 7. (a)Transmission spectra of the cavity for a laser beam with different OAM values. (b) (Top) Reflectance spectrum of the Gaussian mode and locked spectrum. (Bottom) Transmission spectrum of Gaussian laser beam.

Cavity Locking:

A Faraday rotator coupled with a HWP is used to change the *H*-polarized laser beam from the cavity into a *V*-polarized beam, and retains the polarization of the laser beam from coupler 2. The latter beam is reflected towards a power detector by the optical cavity. Error signals generated from the reflectance spectrum is fed back to the piezoelectric-actuated mirror to suppress distance fluctuations between the cavity mirrors; that is, the cavity is locked on the Gaussian mode of the laser beam.

Even though the laser beam from the mode of coupler 2 does not match the cavity very well, the valley of the reflectance spectrum and the peak of the transmission spectrum still have the same frequency. When the cavity is locked on the valley mode, the experimentally measured transmissivity for the Gaussian mode is 90%, and the reflectivity for nonzero OAM mode is higher than 90%. In theory, the transmissivity and reflectivity can be very close to 100%.

## Acknowledgments

This work was supported by the National Fundamental Research Program of China (Grant No. 2 011CBA00200), the National Natural Science Foundation of China (Grant Nos. 11174271, 61275115, 61435011, 61525504).